\documentstyle[12pt]{article}
\newlength{\titleblockheight}
\twocolumn
\leftmargin\leftmargini
\begin{document}
\title{Simulation of the CERN-SPS
        Crystal Extraction Experiment }

\author{\large Valery Biryukov
\\ IHEP Protvino, 142284 Moscow Region, Russia
\\ EPAC 1994 Proceedings, pp.2391-2393
}
\date{}
\maketitle
\abstract{
          Simulation
          of the CERN crystal-extraction experiments at the SPS [1],
	  including the real crystal geometry and SPS parameters,
          as well as multiple turns and passes, has been performed.
	  Two crystals with different geometry have been tried.
	  We argue that in both cases
          the extraction experiment
	  suffered from the crystal edge imperfection.
          For the experimental conditions reproduced,
          the efficiency of the first (''twisted'') crystal
	  was found to be 12--18\% (peak) in the
          angular range of 140--260\,$\mu$rad (FWHM),
          depending on the vertical beam position at the crystal
          location.
	  For the second (''U--shaped'') crystal the peak efficiency
	  was much the same, $\sim$19 \%, while the angular scan
	  FWHM reduced to 70 $\mu$rad.
	  These results, as well as the beam profiles,
	  are in agreement with measurements.
	Finally we suggest some ways to advance this experiment.
}
\setcounter{page}{2391}

\section{Introduction}

The series of crystal extraction experiments is being performed
at CERN\cite{sps}, studying the feasibility of extracting
protons from the halo of the beam circulating in SPS
by a bent silicon crystal.
These studies aim to model the beam parameters expected at the future
Large Hadron Collider,
and have in view possible application of the technique
for a beam extraction from a multi-TeV machine.
Such an extrapolation
requires full understanding of the crystal extraction
process. The present paper investigates
these experiments
by means of a computer simulation.
Our aim is not only to compare theory vs measurement,
but also to discover the factors apparently disturbing
the crystal extraction process in \cite{sps}.

The physics of channeling in a bent crystal is well
established \cite{rc,ufn}.
Beam bending by a crystal is due to the trapping of
some particles in the potential well formed by the field of
 atomic planes, where the particles then  follow
the direction of (are {\em channeling} in) the atomic planes.
The channeling effect survives in a bent crystal until the ratio
of the beam momentum $p$ to the bending radius $R$ becomes
as high as the maximal field gradient ($\sim$6 GeV/cm in silicon).
The crystal bend reduces the phase space available
for channeling, thus decreasing the fraction of particles
channeled.
The scattering processes may cause
the trapped particle to come  to a free state (dechanneling).

In the present simulation we tracked protons
 through the curved crystal lattice
with small, $\sim$1 $\mu$m, steps applying the Monte Carlo code CATCH
\cite{mc}. This code
uses the Lindhard's continuous-potential approach
to the field of atomic planes,
and takes the processes of both single and multiple
scattering on electrons and nuclei into account.
We assumed the crystal to have a perfect lattice.

The simulation of extraction was performed with parameters
matching those of the SPS experiment:
$\alpha_H$\,= 2.07, $\beta_H$\,= 90\,m, $Q_H$\,= 0.635,
$\alpha_V$\,= --0.734, $\beta_V$\,= 24.4\,m, $Q_V$\,= 0.583.
The vertical normalized emittance $\epsilon_V$
was normally 10\,$\pi$\,mm$\cdot$mrad.
Protons were tracked
both in the crystal and in the accelerator for many subsequent
passes and turns before eventual loss either at the
aperture or in interaction with crystal nuclei.
Tracking in the SPS used linear, 4 by 4 transfer matrices.
The crystal was located horizontally at 10\,mm from the beam axis.

The beam was excited by horizontal white noise
at the opposite side of the machine, with kicks of 0.003\,$\mu$rad
r.m.s. value per turn.
The expected impact parameters $b$ and angles $b'$
were studied earlier \cite{herr}.
The $b'$ value there was found to be $\ll \psi_p$.
The $b$ value
should be $\sim$0.7\,$\mu$m,
i.e. quite close to the surface.

The crystal surface quality setting was rather conservative:
miscut angle 200\,$\mu$rad, surface nonflatness 1\,$\mu$m,
plus 1\,$\mu$m thick amorphous layer superposed.
This defines some `septum width' of a few $\mu$m value.
The protons tracking near the surface have taken all these details
(as well as the {\em bent surface}) into account.
We consider here two options. The
{\em first}, with impact parameter below 1\,$\mu$m and
surface setting as described above, excludes the possibility
of channelling in the first pass through the crystal.
This is compared to the {\em second} option, when
the crystal surface is assumed perfect, i.e. with zero septum
width.

\section{''Twisted'' crystal}

The first extraction experiments \cite{sps}
were performed with a ''Serpukhov-type'' bending device,
and have given results more sophisticated than expected:
the extraction efficiency was rather small, around 10\%;
extraction occurred in the angular range of 200\,$\mu$rad FWHM,
much wider than the beam divergence or critical channelling angle
(14 $\mu$rad);
two peaks appeared both in the horizontal and vertical profile
of the extracted beam at the
crystal orientation far away from the best one;
the angular range of extraction depended on the vertical beam position.

In the laser measurements it was found that the bent crystal
is deformed vertically, called `twist' in the following.
As a result, the planar direction at
the entrance to the crystal
was a function of the vertical position.
The measured crystal properties
were used as input to the
simulations.
The total bending found
for this (3 cm long, 1.5 mm thick) crystal was 8.5\,mrad.
\begin{table}[htb]
\caption{Summary of simulation for $\epsilon_V$=10$\pi$ and 1.5$\pi$.}
\label {tab2}
\begin{center}
\begin{tabular}{|c|c|c|}
\hline
 & \multicolumn{2}{|c|}{\large \sc Simulation}
  \\
FWHM & Bad surface  & Good surface   \\
\hline \hline
{\bf nominal}  & &  \\
X (mm) & 1.9 & 1.7   \\
Y (mm) &  3.4 & 1.7  \\
scan ($\mu$rad) &  140 & 30 \\
\hline
efficiency(\%) & 18 & 40  \\
\hline \hline
{\bf bump 1\,mm}  & & \\
X (mm) & 3.5 & 1.1   \\
Y (mm) & 3.1 & 1.8     \\
scan ($\mu$rad) & 170 & 90  \\
\hline
efficiency(\%) & 16 & 24 \\
\hline \hline
{\bf bump 2\,mm}  & & \\
X (mm) & 2.2  & 1.0    \\
Y (mm) & 2.2  & 2.0    \\
scan ($\mu$rad) & 260 & 200  \\
\hline
efficiency(\%) & 12 & 15 \\
\hline  \hline
{\bf $\epsilon_V$=1.5} & & \\
X (mm) & 1.6 & 0.8   \\
Y (mm) & 1.7 & 0.7   \\
scan ($\mu$rad) & 110 & 25  \\
\hline
efficiency(\%) & 19 & 52 \\
\hline
\end{tabular}
\end{center}
\end{table}
In the experiment, different vertical beam positions (bump)
and different vertical
emittances $\epsilon_V$ have been tried in order
to investigate the twist effect.
    Our simulation followed the program of the SPS experiment.
    Table~1 summarizes the extraction efficiencies,
 the widths of the angular scan and of the extracted-beam
 profiles (observed 20 m downstream)
 obtained in the simulation for $\epsilon_V=10\pi$
 and 1.5$\pi$ mm$\cdot$mrad.
\begin{table}[htb]
\caption{
The peak efficiency (\%) plotted vs
$\epsilon_V$  and the vertical beam position (bump).
}
\label {tab3}
\begin{center}
\begin{tabular}{|c|ccc|}
\hline
 & & &  \\
 Bump &  & {\large {$\epsilon_V$}} [$\pi\cdot$mm$\cdot$mrad] & \\
 {[mm]} & $\;$ 1.5  &  10  &  20  $\;$  \\
$\Downarrow$  & & & \\
\hline
 & & & \\
 0  & 19  & 18  & 16  \\
&  &  &  \\
 1  &  & 16  & 15  \\
&   &  &   \\
 2  &  & 12 & \\
& & & \\
\hline
\end{tabular}
\end{center}
\end{table}
The extraction efficiency as a function of tilt angle
is shown in Fig.\,1 both
for nominal position ($\bullet$) and bump of 2 mm (o).
The increase of $\epsilon_V$ to 20 $\pi$\,mm$\cdot$mrad
has slightly reduced the efficiencies.
This is summarized in Table~2, where the peak efficiency
is plotted as a function of $\epsilon_V$
and the vertical position of the beam at crystal.

In the SPS experiment, the nominal vertical position
was $\sim$1\,mm away from the `twist middle'.
The $\epsilon_V$ typical value was
$\sim$10\,$\pi$\,mm$\cdot$mrad. Therefore the number of 16\% for the peak
efficiency (and 170\,$\mu$rad for the angular scan FWHM)
seems the most relevant one to compare with the experimental results,
namely 10$\pm$1.7\% and 200\,$\mu$rad.
The systematic error caused by some uncertainties in the vertical
position, $\epsilon_V$ value, and twist detail,
can be estimated from Table~2 as a few per cent for the peak
efficiency.
We may conclude that the perfect-surface simulation has often predicted
narrow high peaks, which have not been observed.
The imperfect-surface option roughly followed the
experimental observation.
Both the angular scan FWHM
and the efficiency value depend on the bump
in the same way as it was observed at the SPS.
The extracted beam vertical width tends to decrease
with vertical displacement, while the horizontal width
has no trend but sizeable fluctuations.
Both features seem to agree with experimental observation.
There is  quantitative agreement with the measurement,
but the price for this agreement is  the assumption
of  crystal edge imperfection.
The twist effect was minor for the {\em peak} efficiency,
and cannot alone explain the large angular width of the extraction.
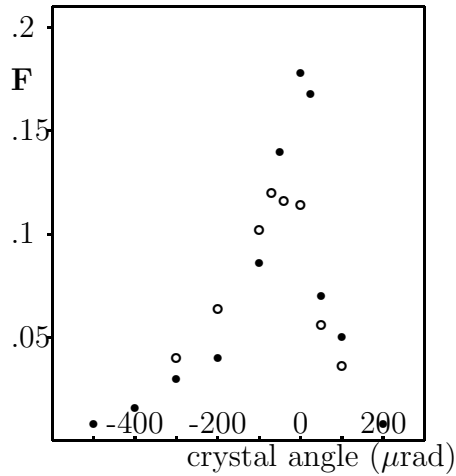
\begin{figure}
\begin{center}
\setlength{\unitlength}{.55mm}
\begin{picture}(120,105)(-70,-4)
\thicklines

\put(0,57){\circle{2}}
\put(5,28){\circle{2}}
\put(10,18){\circle{2}}
\put(-4,58){\circle{2}}
\put(-7,60){\circle{2}}
\put(-10,51){\circle{2}}
\put(-20,32){\circle{2}}
\put(-30,20){\circle{2}}

\put(0,89){\circle*{2}}
\put(2.5,84){\circle*{2}}
\put(5,35){\circle*{2}}
\put(10,25){\circle*{2}}
\put(20,4){\circle*{2}}
\put(-5,70){\circle*{2}}
\put(-10,43){\circle*{2}}
\put(-20,20){\circle*{2}}
\put(-30,15){\circle*{2}}
\put(-40,8){\circle*{2}}
\put(-50,4){\circle*{2}}

\put(-60,0) {\line(1,0){90}}
\put(-60,0) {\line(0,1){105}}
\put(-60,105) {\line(1,0){90}}
\put(30,0){\line(0,1){105}}
\multiput(-50,0)(10,0){8}{\line(0,1){1}}
\put(-.5,2.){\makebox(1,.5)[b]{0}}
\put(19.5,2.){\makebox(1,.5)[b]{200}}
\put(-20.5,2.){\makebox(1,.5)[b]{-200}}
\put(-40.5,2.){\makebox(1,.5)[b]{-400}}
\multiput(-60,25)(0,25){4}{\line(-1,0){1}}
\put(-70,100){\makebox(1,.5)[l]{.2}}
\put(-70,75){\makebox(1,.5)[l]{.15}}
\put(-70,50){\makebox(1,.5)[l]{.1}}
\put(-70,25){\makebox(1,.5)[l]{.05}}

\put(-70,85){\bf F}
\put(-30,-6){ crystal angle ($\mu$rad)}

\end{picture}
\caption{ Angular scans of efficiency
for twisted crystal. Nominal position ($\bullet$) and bump of 2 mm (o)}
\end{center}
\end{figure}

\section{''U-shaped'' crystal }

After the above analysis, a prediction was made in \cite{sl}
for the new, ''U-shaped'' crystal\cite{rd2} which has no twist.
This crystal is 4 cm long with 8.2 mrad bending.
Fig.~2 shows the angular scan simulated for this crystal
with edge imperfection,
which is in good agreement with the first experimental results\cite{rd2}.
As compared to the ''twisted'' crystal, there was expected no
sizeable change of efficiency, in accord with the first observation.
Such an agreement indicates that the new crystal also has some edge
imperfection.
For an ideal crystal and a very parallel incident beam
the simulation gives the peak efficiency $\sim$50\% and
the narrow angular scan (25\,$\mu$rad FWHM).
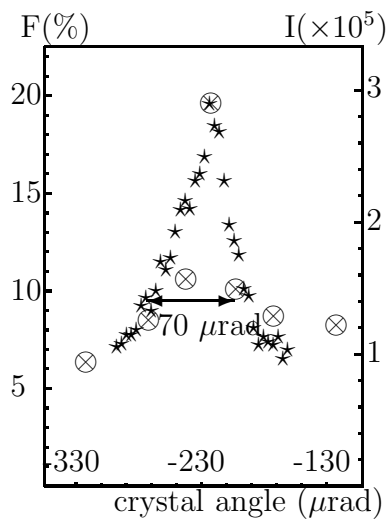
\begin{figure}
\begin{center}
\setlength{\unitlength}{.65mm}
\begin{picture}(70,100)(-6,-20)
\thicklines

\put(0,-19) {\line(1,0){65}}
\put(0,-19) {\line(0,1){90}}
\put(0,71) {\line(1,0){65}}
\put(65,-19) {\line(0,1){90}}
\multiput(0,-19)(0,3.994){23}{\line(1,0){.6}}
\multiput(0,-19)(0,19.97){5}{\line(1,0){1.2}}
\put(-7,0.99){\makebox(2,1)[l]{5}}
\put(-7,20.6){\makebox(2,1)[l]{10}}
\put(-7,40.83){\makebox(2,1)[l]{15}}
\put(-7,60.8){\makebox(2,1)[l]{20}}
\multiput(65,-19)(0,5.4){16}{\line(-1,0){.8}}
\multiput(65,-19)(0,27){4}{\line(-1,0){1.6}}
\put(67,8){\makebox(2,1)[r]{1}}
\put(67,35){\makebox(2,1)[r]{2}}
\put(67,62.2){\makebox(2,1)[r]{3}}

\multiput(6.36,-19)(5.128,0){10}{\line(0,1){.4}}
\multiput(6.36,-19)(25.64,0){3}{\line(0,1){.7}}
\put(30,-16){\makebox(2,1)[b]{-230}}
\put(4.36,-16){\makebox(2,1)[b]{-330}}
\put(55.64,-16){\makebox(2,1)[b]{-130}}

\put(-5,73){F(\%)}
\put(49,73){I($\times 10^5$)}
\put(12,-24){ crystal angle ($\mu$rad)}

\put( 3.69, 4.8){ $\otimes$}
\put(16.51,13.3){ $\otimes$}
\put(24.20,21.7){ $\otimes$}
\put(29.33,57.7){ $\otimes$}
\put(34.46,19.7){ $\otimes$}
\put(42.15,14.2){ $\otimes$}
\put(54.97,12.3){ $\otimes$}

\put(23,12){70 $\mu$rad}
\put(30,19){\vector(-1,0){9.2}}
\put(30,19){\vector(1,0){8.75}}

\put(11,8){ $\star$}
\put(12,8.7){ $\star$}
\put(13,10.5){ $\star$}
\put(14,10.5){ $\star$}
\put(15,11.5){ $\star$}
\put(16,16.4){ $\star$}
\put(17,18){ $\star$}
\put(18,15.4){ $\star$}
\put(19,19.5){ $\star$}
\put(20,25.5){ $\star$}
\put(21,23.7){ $\star$}
\put(22,26.3){ $\star$}
\put(23,31.5){ $\star$}
\put(24,36){ $\star$}
\put(25,38){ $\star$}
\put(26,36.3){ $\star$}
\put(27,42){ $\star$}
\put(28,43.5){ $\star$}
\put(29,47){ $\star$}
\put(30,57.7){ $\star$}
\put(31,53.3){ $\star$}
\put(32,52){ $\star$}
\put(33,42.1){ $\star$}
\put(34,33){ $\star$}
\put(35,29.7){ $\star$}
\put(36,26.8){ $\star$}
\put(37,19.9){ $\star$}
\put(38,18.5){ $\star$}
\put(39,12){ $\star$}
\put(40,8.5){ $\star$}
\put(41,10){ $\star$}
\put(42,9){ $\star$}
\put(43,8.5){ $\star$}
\put(44,10){ $\star$}
\put(45,5.5){ $\star$}
\put(46,7.4){ $\star$}

\end{picture}
 \caption
  { Prediction ($\otimes$, left scale)
and measurement ($\star$, right scale) for the U-shaped crystal.}
\end{center}
\end{figure}
\section{Experiment Advancement  }
An essential advancement of the crystal extraction experiment would be
the measurement of the first-pass contribution to efficiency.
This is quite important for understanding both of the crystal-edge work
(measuring ''septum width'') and of the crystal co-work with
the other elements of accelerator (multiple passes/turns).
One can distinguish the first pass from the secondary ones owing to
their difference in time (first comes first) and space
(secondary pass comes deeper in the crystal).
The time-synchronyzing (''kick mode'') has difficulty
caused by unavoidable mixing of the ''first-pass'' protons
from later turns.
For a spatial separation
one way is a thin amorphous layer superposed on the crystal
surface to suppress the first pass (W.Herr).
Here are some other ways:

{\bf Microscope.} Applying a skew cut on the exit face
of a crystal, one can make the bending angle to depend on the impact
parameter $b$ (a principle of any microscope). The bent beam
profile downstream the crystal would be then an amplified
profile of this beam at the crystal face.

{\bf Energy loss \cite{dedz}.} The  $\delta E/\delta z$ {\em deposited}
in the crystal may indicate whether the proton is extracted with the
first pass or with the secondary one, due to the escape of
$\delta$-electrons near the surface (depending on $b$).
And vice versa, selecting 'first-pass' in some other way one may
observe an interesting $\delta E/\delta z$ spectrum
(fwhm$\approx$0).

{\bf Masking \cite{sl}.} One can scrape the circulating beam by a
collimator during the crystal extraction, thus suppressing the secondary
passes.

{\bf Tune resonance \cite{sl}.} It was shown that for an {\em imperfect}
edge of a crystal the extraction efficiency drops near the tune resonance
(for the same given parameters of incident particles).
The resonance both dip and width depend on
(are roughly proportional to) the septum width.

{\bf Crystal rotation in ''FNAL geometry'' \cite{91}.}
In the different geometry of extraction \cite{e853}
there is an extra degree of freedom. The crystal rotation {\em normal}
to bending plane results in a controlled change (with steps
of $\sim$0.02 $\mu$m) of septum width.

Fig.~3 shows the extraction efficiency as function of L,
simulated for an untwisted crystal with constant curvature.
A short ($\sim$1 cm here) length could be a ''cure'' to any
imperfection of a crystal, as the protons make more passes.
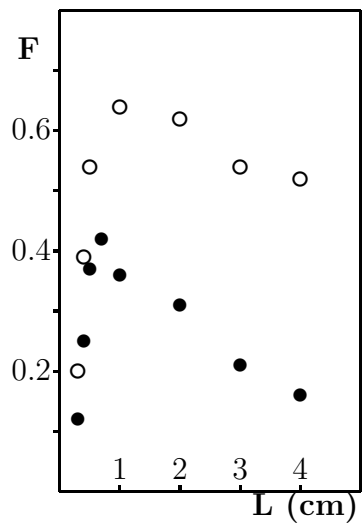
\begin{figure}
\begin{center}
\setlength{\unitlength}{.8mm}
\begin{picture}(80,89)(-10,-1)
\thicklines

\put(3,20){\circle{2}}
\put(4,39){\circle{2}}
\put(5,54){\circle{2}}
\put(10,64){\circle{2}}
\put(20,62){\circle{2}}
\put(30,54){\circle{2}}
\put(40,52){\circle{2}}

\put(3,12){\circle*{2}}
\put(4,25){\circle*{2}}
\put(5,37){\circle*{2}}
\put(7,42){\circle*{2}}
\put(10,36){\circle*{2}}
\put(20,31){\circle*{2}}
\put(30,21){\circle*{2}}
\put(40,16){\circle*{2}}

\put(0,0) {\line(1,0){50}}
\put(0,0) {\line(0,1){80}}
\put(0,80) {\line(1,0){50}}
\put(50,0){\line(0,1){80}}
\multiput(10,0)(10,0){5}{\line(0,1){1}}
\put(9.5,2.){\makebox(1,.5)[b]{1}}
\put(19.5,2.){\makebox(1,.5)[b]{2}}
\put(29.5,2.){\makebox(1,.5)[b]{3}}
\put(39.5,2.){\makebox(1,.5)[b]{4}}
\multiput(0,10)(0,10){7}{\line(-1,0){1}}
\put(-8,20){\makebox(1,.5)[l]{0.2}}
\put(-8,40){\makebox(1,.5)[l]{0.4}}
\put(-8,60){\makebox(1,.5)[l]{0.6}}

\put(-7,72){\bf F}
\put(32,-4){\bf L (cm)}

\end{picture}
\caption{ Extraction efficiency vs crystal length.
For edge imperfection ($\bullet$) and ideal crystal (o).}
\end{center}
\end{figure}

\end{document}